\documentclass[12pt,]{article}
\usepackage{url}

\providecommand{\tightlist}{%
  \setlength{\itemsep}{0pt}\setlength{\parskip}{0pt}}

\usepackage{lmodern}
\usepackage{amssymb,amsmath}
\usepackage{ifxetex,ifluatex}
\usepackage{fixltx2e} % provides \textsubscript
\ifnum 0\ifxetex 1\fi\ifluatex 1\fi=0 % if pdftex
  \usepackage[T1]{fontenc}
  \usepackage[utf8]{inputenc}
\else % if luatex or xelatex
  \ifxetex
    \usepackage{mathspec}
    \usepackage{xltxtra,xunicode}
  \else
    \usepackage{fontspec}
  \fi
  \defaultfontfeatures{Mapping=tex-text,Scale=MatchLowercase}
  
\fi
\IfFileExists{upquote.sty}{\usepackage{upquote}}{}
\IfFileExists{microtype.sty}{%
\usepackage{microtype}
\UseMicrotypeSet[protrusion]{basicmath} % disable protrusion for tt fonts
}{}
\usepackage{color}
\usepackage{fancyvrb}

\DefineVerbatimEnvironment{Highlighting}{Verbatim}{commandchars=\\\{\}}
\usepackage{framed}
\definecolor{shadecolor}{RGB}{248,248,248}
\newenvironment{Shaded}{\begin{snugshade}}{\end{snugshade}}
\newcommand{\KeywordTok}[1]{\textcolor[rgb]{0.13,0.29,0.53}{\textbf{{#1}}}}
\newcommand{\DataTypeTok}[1]{\textcolor[rgb]{0.13,0.29,0.53}{{#1}}}
\newcommand{\DecValTok}[1]{\textcolor[rgb]{0.00,0.00,0.81}{{#1}}}

\newcommand{\FloatTok}[1]{\textcolor[rgb]{0.00,0.00,0.81}{{#1}}}

\newcommand{\CharTok}[1]{\textcolor[rgb]{0.31,0.60,0.02}{{#1}}}

\newcommand{\StringTok}[1]{\textcolor[rgb]{0.31,0.60,0.02}{{#1}}}

\newcommand{\OtherTok}[1]{\textcolor[rgb]{0.56,0.35,0.01}{{#1}}}

\newcommand{\NormalTok}[1]{{#1}}
\ifxetex
  \usepackage[setpagesize=false, % page size defined by xetex
              unicode=false, % unicode breaks when used with xetex
              xetex]{hyperref}
\else
  \usepackage[unicode=true]{hyperref}
\fi
\hypersetup{breaklinks=true,
            bookmarks=true,
            pdfauthor={},
            pdftitle={Designing Modular Software: A Case Study in Introductory Statistics},
            colorlinks=true,
            citecolor=blue,
            urlcolor=blue,
            linkcolor=magenta,
            pdfborder={0 0 0}}
\title{\bf Designing Modular Software: A Case Study in Introductory Statistics}
\author{Eric Hare \\ Iowa State University \\ \texttt{}  and \\ Andee Kaplan \\ Iowa State University \\ \texttt{} }
\date{}

\usepackage{graphicx}
\usepackage{multirow}
\usepackage{color}
\usepackage{float}

\bigskip

\begin{document}

\def\spacingset#1{\renewcommand{\baselinestretch}%
{#1}\small\normalsize} \spacingset{1}

\maketitle

\begin{abstract}
Modular programming is a development paradigm that emphasizes
self-contained, flexible, and independent pieces of functionality. This
practice allows new features to be seamlessly added when desired, and
unwanted features to be removed, thus simplifying the user-facing view
of the software. The recent rise of web-based software applications has
presented new challenges for designing an extensible, modular software
system. In this paper, we outline a framework for designing such a
system, with a focus on reproducibility of the results. We present as a
case study a Shiny-based web application called \texttt{intRo}, that
allows the user to perform basic data analyses and statistical routines.
Finally, we highlight some challenges we encountered, and how to address
them, when combining modular programming concepts with reactive
programming as used by Shiny.
\end{abstract}

\noindent%
\vfill

\newpage
\spacingset{1.45} % DON'T change the spacing!

\section{Background}\label{background}

Modularity is a pervasive concept in computer science, extending from
the design of systems (Parnas 1972), to the design of software
(Szyperski 1996). Modularity offers several advantages to both a
developer and a user. In particular, functionality can be dynamically
loaded and unloaded depending on the particular use case. Open source
modular software precipitates the possibility of extensions contributed
by a wide array of programmers, which can allow the software to morph
into areas that weren't anticipated early in development. In the
statistics realm, R (R Core Team 2014) is a prime example of the virtues
of modular programming. As of this writing, The Comprehensive R Archive
Network (CRAN) contains over 9000 source packages which can be installed
and dynamically loaded in a particular session as needed.

Other statistics software also makes use of a number of these ideas.
Microsoft Excel and JMP both include support for extensions, called
macros and add-ins respectively, which allow programming routines to be
written extending the base functionality of these programs. Compared
with R, however, these programs don't maintain a large central
repository of publicly available extensions on the level of CRAN. There
are also software packages building upon R, and thus gaining the
advantages of CRAN natively, such as R Commander (Fox 2005) and Deducer
(Fellows 2012), which each provide a graphical front-end to many
statistical functions in R. One thing these software packages all have
in common is the requirement of local installation and configuration,
which means certain operating systems and platforms will not support
their use.

With the advantages of R clear, an approach to building statistical
software and statistical learning tools would be to attempt to generate
interest in programming, which could help naturally ease the transition
into the use of R. Multiple software packages have recently been written
in an attempt to spur this interest in R programming and statistics.
DataCamp's (DataCamp 2014) courses are a user-friendly way to learn
basic R programming and data analysis techniques. Swirl (Carchedi et al.
2014) is a similar interactive tool to make learning R more fun by
learning it within R itself. Project MOSAIC (Pruim, Kaplan, and Horton
2014) has created a suite of tools to simplify the teaching of
statistics in the form of an R package. The primary goal of DataCamp and
Swirl is to teach R programming, rather than facilitate the learning of
introductory statistics.

Modern web technologies have enabled a new generation of software
packages that reside solely on the web, which eliminates the issue of
local installation and helps abstract away some of the more challenging
programming aspects of working directly with R. Upon the release of
RStudio's Shiny (RStudio and Inc. 2014) it became easier for an R-based
analysis to be converted to an interactive web application. Several
recent software packages have built upon Shiny to provide a web-based
system based on R. One such package is iNZight Lite (Wild 2015) which
attempts to expose students to data analysis without requiring
programming knowledge. Like most web-based systems, this does not
include reproducible R code, which limits its usefulness in a scientific
or academic setting. Another package is called Radiant (Nijs 2016),
which is a web-based application with the aim of furthering business
education and financial analysis. While the application is modular and
extensible, it does require installation and hosting and is inundated
with more features than necessary for an introductory student. An
overview of the comparison between the features of these statistical
software packages is presented in Table \ref{tab:compare}. Partial
fulfillment of requirements is noted in the table, as well as a measure
of the complexity of functionality offered by default. For example,
\texttt{R} does have an associated Graphical User Interface (GUI),
however this interface is very limited, thus only partially fulfilling
the behavior of a GUI.

\begin{table}[ht]
\centering
\begin{tabular}{llllllll}
  \hline
Software & GUI & Install & Modular & Web & Extensible & Reproducible & Features \\ 
  \hline
intRo & Yes & No & Yes & Yes & Yes & Yes & Limited \\ 
  JMP & Yes & Yes & Partial & No & Partial & No & Full \\ 
  R & Partial & Yes & Yes & No & Yes & Yes & Full \\ 
  Rcmdr & Yes & Yes & No & No & Yes & Yes & Moderate \\ 
  Deducer & Yes & Yes & No & No & Yes & Yes & Moderate \\ 
  MOSAIC & No & Yes & No & No & No & Yes & Limited \\ 
  iNZight Lite & Yes & No & Yes & Yes & Partial & No & Limited \\ 
  Radiant & Yes & Yes & Yes & Yes & Partial & Yes & Moderate \\ 
   \hline
\end{tabular}
\caption{A comparison of statistical software packages across the metrics of usability, modularity, extensibility, and reproducibility. Partial fulfillment of requirements is noted in the table, as well as a measure of the complexity of functionality offered by default.} 
\label{tab:compare}
\end{table}

Though challenging in a GUI, a reproducibility framework has three key
advantages. First, it eases a student who may be intimidated by
programming into the idea that interacting with a user interface is
really just a frontend for code. Seeing the correspondence between
graphical clicks and printed code should help lessen the fear of coding
that many students may have. Second, an analysis created by a
reproducible software system can be brought in an R session to easily
assess and extend the results. Finally, with the help of knitr (Xie
2015) and rmarkdown (Allaire et al. 2014), ``printing'' the results of a
reproducible software system analysis amounts to nothing more than
executing the R code on the server, adding another layer of
reproducibility. These concepts are important because they encourage
best practices with regards to disclosure of analysis methods in
research (Baggerly and Berry 2011; Xie 2015).

Based on the above, we believe a modern software system should be
\textbf{modular}, \textbf{extensible}, \textbf{web-based}, and foster
\textbf{reproducibility}. We have developed a case-study application
called \texttt{intRo} which we will use to illustrate our method of
developing a system meeting these criteria. The paper is structured as
follows: Section \ref{case-study-intro} introduces the application, its
features and its usability, and provides motivations for why it was
built. Section \ref{intro-design-decisions} provides technical details
on how we built \texttt{intRo}, by walking through the underlying
modularity, reproducibility, and reactive framework, as well as how it
can be used to develop other software systems with these properties.
Finally, Section \ref{conclusions-and-future-work} discusses some future
possibilities and limitations of both \texttt{intRo}, and modular
systems in general.

\section{\texorpdfstring{Case Study:
\texttt{intRo}}{Case Study: intRo}}\label{case-study-intro}

The widespread adoption of R as a tool for statistical analysis has
undoubtedly been an important development for the scientific community.
However, using R in most cases still requires a basic knowledge of
programming concepts which may pose a steep learning curve for the
introductory statistics student (Tan, Ting, and Ling 2009). This
additional time commitment may explain why introductory courses often
utilize point-and-click applications, even if the instructor
himself/herself uses R in their own work. Still, some compromises must
be made when using many graphical applications, including dealing with
software licenses and unsupported desktop platforms. From the
instructor's perspective in particular, managing a large group of
software licenses for students with various computing environments and
versions could wind up being extremely cumbersome.

In teaching Introduction to Business Statistics at Iowa State
University, we witnessed profound struggles by students attempting to
practice introductory concepts discussed in class using current
software. Scrimshaw (2001) notes in his manuscript that ``open-ended
packages, like any others, may create obstacles to learning simply
through their lack of user-friendliness in the sheer mechanics of
operating them, rather than any intrinsic difficulty in the
content\ldots{}.'' In our own experience teaching, students' struggles
were often directly related to the use of the software and not any sort
of fundamental misunderstanding of the material, in agreement with
Scrimshaw's finding.

These challenges led us to create an introductory statistics application
which we call \texttt{intRo}, available at
\url{http://www.intro-stats.com}. \texttt{intRo} offers a number of key
advantages over traditional statistics software, including ease of
access and an aim to foster student interest in coding. Attempting to
entirely hide the programming aspect from students, even in introductory
classes, is a lost opportunity to get students interested in statistical
computing. It is also a lost opportunity reaching students who learn
differently or have a computational background. Another advantage is its
modular structure, which allows course instructors to tailor the
application towards the needs of a particular class, rather than accept
a piece of software as is. Additionally, \texttt{intRo} stands apart
from new tools in that it is a supplement to an existing class, fully
usable by a beginning statistics student. An accompanying R package,
titled \texttt{intRo} and available on GitHub, assists in the
downloading, running, and deploying of \texttt{intRo} instances (See
Section \ref{deploying-intro-instances}).

Three fundamental philosophies that guided the creation of
\texttt{intRo}. In particular, \texttt{intRo} is \emph{easy} to use and
can be an \emph{exciting} part of learning statistics. Additionally,
\texttt{intRo} is an \emph{extensible} tool, allowing for a course
instructor using \texttt{intRo} to tailor the tool for his or her own
classroom needs.

\begin{figure}[H]
\centering
\includegraphics[width=\linewidth]{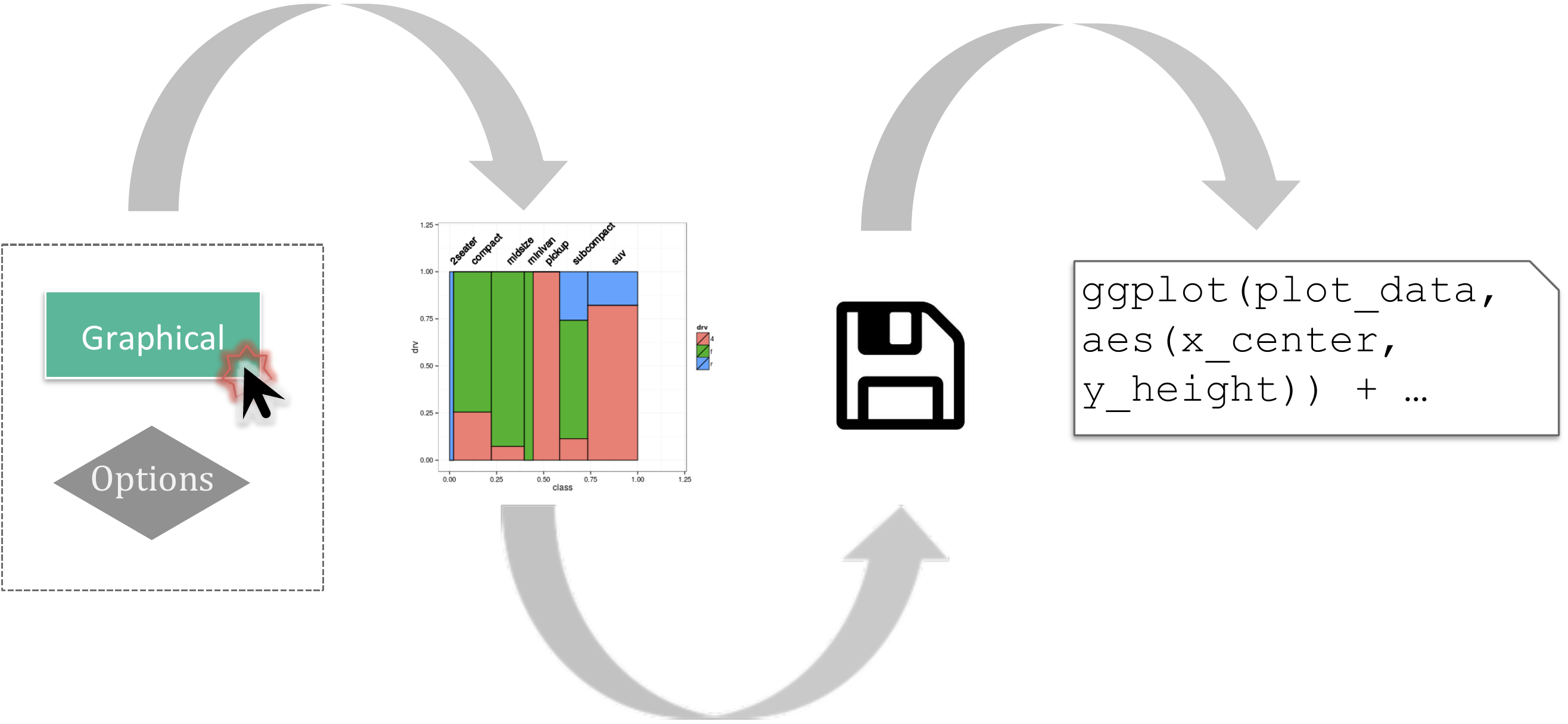}
\caption{A schematic of the typical student experience of generating a result in \texttt{intRo}. In this instance, a student clicks on the Graphical tab to create a mosaic plot. The student sees the plot, and elects to click the save button to store the plot (and its corresponding code) to the final compendium.}
\label{fig:user_experience}
\end{figure}

In the development of \texttt{intRo}, we focused on aspects of the user
interface (UI) and output that make it easy to pick up without extensive
training. We used large, easy to click icons in the page header to help
students find what they need more easily. We also made the functionality
available the minimal necessary for an introductory statistics course.
Figure \ref{fig:user_experience} presents a schematic of the simple
steps a student takes to generate a result in \texttt{intRo}. In this
instance, a student clicks on the Graphical tab to create a mosaic plot.
The student sees the plot, and elects to click the save button to store
the plot and its corresponding code to the final compendium.

\begin{figure*}[ht!]
\centering
\includegraphics[width=\linewidth]{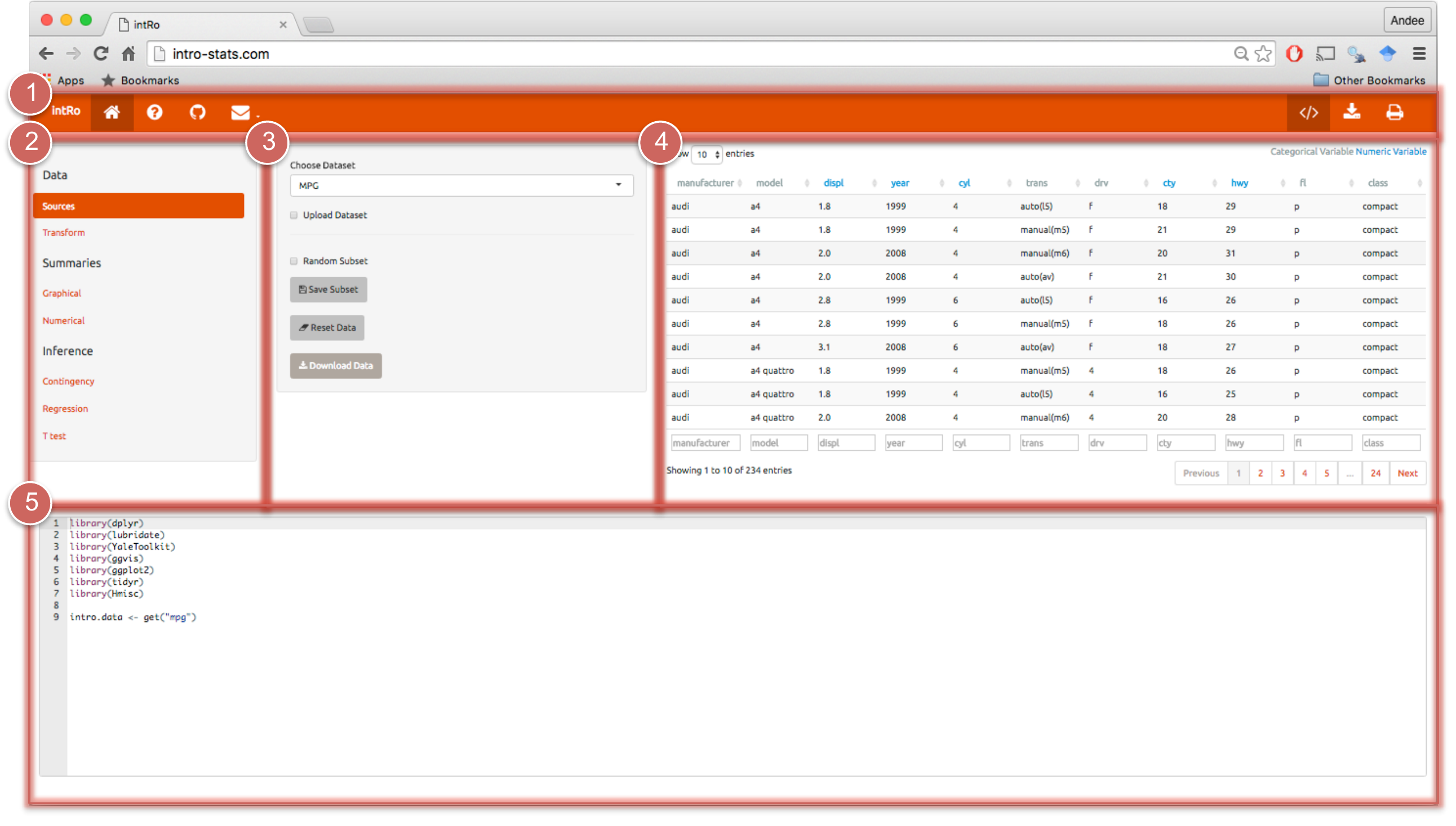}
\caption{The five elements that comprise the \texttt{intRo} application: 1) top navigation, 2) side navigation, 3) options panel, 4) results panel, and 5) code panel.}
\label{fig:ui}
\end{figure*}

Beyond being simple, \texttt{intRo} is also consistent. The tool is
organized around specific tasks a student may perform in the process of
a data analysis, called modules. To the student, a module is simply a
page of statistics functionality that maintains a consistent layout,
helping the student to become familiar with the location of the options,
the results, and the code. Figure \ref{fig:ui} highlights the five
elements that comprise the \texttt{intRo} interface.

\begin{enumerate}
\def\labelenumi{\arabic{enumi}.}
\tightlist
\item
  \textbf{Top Navigation} - The top navigation bar includes two sets of
  clickable icons. The left-aligned buttons are informational buttons.
  The first is a link to \texttt{intRo}. The second is a link to the
  documentation page. The third is a link to the GitHub repository where
  the code for \texttt{intRo} is housed. The final button is a link to
  our websites, which contain contact information if there are any
  questions or comments. The right-aligned buttons are \texttt{intRo}
  utilities. The first is a link to toggle the visibility of the code
  panel (5). The middle icon downloads an rmarkdown document of the
  analysis performed. The last is a link to print the stored module
  results, and the associated code (if visible).
\item
  \textbf{Side Navigation} - The side navigation panel includes a list
  of data analysis tasks.
\item
  \textbf{Options Panel} - The options panel includes task-specific
  options which the student can use to customize their results.
\item
  \textbf{Results Panel} - The results pane displays the result of the
  selected module and options.
\item
  \textbf{Code Panel} - The code panel displays the R code used to
  generate the results from the student's \texttt{intRo} session. The
  code panel is shown by default to facilitate a transition to coding,
  but can be hidden by clicking the code toggle button in the Top
  Navigation bar.
\end{enumerate}

The modules included in \texttt{intRo} are split into three higher level
categories - data, summaries, and inference. Under each of these
categories, there are seven default modules, which perform specific data
analysis tasks that employ an easy to use point-and-click interface.
More modules can easily be added by an instructor, as detailed in
Section \ref{designing-for-modularity}. The default modules support
uploading and downloading a dataset, transforming variables, graphical
and numerical summaries, simple linear regression, contingency tables,
and T-tests.

\texttt{intRo} has an ulterior motive as well: to get students excited
about programming. By navigating about the user interface of
\texttt{intRo}, students are actually creating a fully-executable,
reproducible R script that they can download and run locally as well as
viewing the script change real-time within the application. This code
creation element of \texttt{intRo} is meant to generate excitement about
programming in R and empower students to feel that they can generate
code as well. \texttt{intRo} uses rmarkdown's render function in order
to print the results, by dymanically executing the student's R script.
By default, the output will include the R code, but if the student
elects to hide the source code by clicking the code toggle button at the
top, the code will not appear in the printed results.

On the front end, user interaction with \texttt{intRo} is split into
bitesize chunks that we call modules. In \texttt{intRo}'s context,
modules are self-contained pieces of functionality which implement
common statistical procedures. These modules form the core functionality
of \texttt{intRo} and are discussed at length in the next section.

\section{\texorpdfstring{\texttt{intRo} Design
Decisions}{intRo Design Decisions}}\label{intro-design-decisions}

In this section, we detail the design choices surrounding
\texttt{intRo}'s extensibility. We have designed it in such a way that
these ideas can be used in other Shiny-based software projects.

\subsection{Designing for Modularity}\label{designing-for-modularity}

\begin{figure*}[ht!]
\centering
\includegraphics[width=\linewidth]{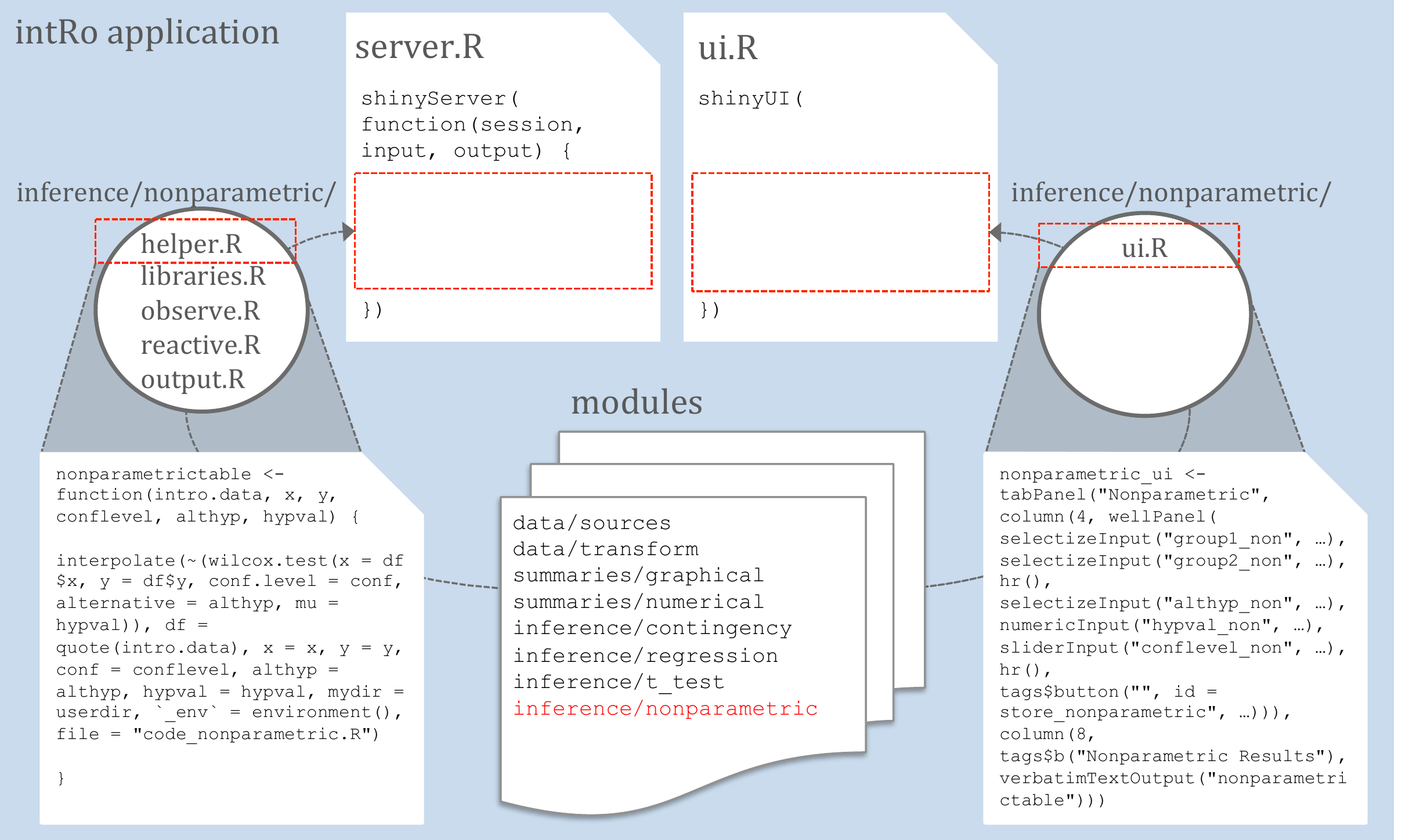}
\caption{This figure depicts how the Shiny \texttt{server.R} and \texttt{ui.R} files are populated using the modular structure within \texttt{intRo}. \texttt{intRo} modules are split up into multiple R scripts which are included either in Shiny's user interface or server definitions. At runtime, the \texttt{intRo} sources in the specified modules (contained in the modules folder) to dynamically generate the functionality available in the application. This allows for the specific functionality needed to be determined and adjusted by the individual course instructor. In this example, the instructor is electing to include a nonparametric module, which is not enabled by default, to allow the students to perform a wilcoxon rank sum test.}
\label{fig:app_creation_modules}
\end{figure*}

An \texttt{intRo} module is a set of self-contained executible R scripts
that together produce a set of introductory statistics functionality.
\texttt{intRo} modules were designed in this way to allow for simple
dynamic creation of the user interface at run-time, as well as ease the
process of converting existing analysis code to the \texttt{intRo}
framework. A high-level diagram of this process is given in Figure
\ref{fig:app_creation_modules}. \texttt{intRo} modules are split up into
multiple R scripts which are included either in Shiny's user interface
or server definitions. At runtime, the \texttt{intRo} sources in the
specified modules (contained in the modules folder) to dynamically
generate the functionality available in the application. This allows for
the specific functionality needed to be determined and adjusted by the
individual course instructor. In this example, the instructor is
electing to include a nonparametric module, which is not enabled by
default, to allow the students to perform a wilcoxon rank sum test.

Section \ref{dynamic-ui-generation} provides some technical details on
how we implemented this. For the rest of this section, we focus on the
structure and development of the modules themselves, to aid in the
process of creating and deploying new modules.

Modularity was a design decision we focused on from the start of
\texttt{intRo}'s development. There are some practical benefits to
thinking of related statistics and data science functionality in terms
of modules. Because modules are enabled at run-time, including new
functionality is as simple as downloading and placing a module within
\texttt{intRo}'s modules folder, or removing existing modules from that
folder. Furthermore, errors can be more easily isolated to specific
components. For instance, if an error is encountered, simply disabling
the module can provide a temporary workaround while the issue is
identified. Finally, modularity helps to organize the different pieces
of code into functionality chunks that make it easier for developers to
maintain.

\texttt{intRo} modules are not to be confused with Shiny modules (Cheng
2015). Shiny modules are a recent feature added to Shiny which allows
the bundling of inputs and outputs into a single set of functionality.
They are more general and suitable for any application. \texttt{intRo}
modules are for statistics functionality and work within the
\texttt{intRo} application only.

An \texttt{intRo} module consists of the following scripts:

\begin{itemize}
\tightlist
\item
  \emph{helper.R} - R code that performs some statistical analysis or
  transformation. This would typically be in the form of a function, and
  similar to any standard R script.
\item
  \emph{libraries.R} - Code to load any libraries which are not part of
  core R.
\item
  \emph{observe.R} - Shiny observer code typically used to update
  choices of an input box.
\item
  \emph{output.R} - Shiny output code defining the results of the
  analysis that should be displayed to the student.
\item
  \emph{reactive.R} - Shiny reactives, typically containing data that
  depend on inputs.
\item
  \emph{ui.R} - Shiny user interface definition, including the placement
  of the inputs and outputs.
\end{itemize}

The modules provided with \texttt{intRo} are contained in the modules
folder. The top level directory in the modules folder defines the
category of the module (currently \texttt{data}, \texttt{summaries}, or
\texttt{inference}). Within each of these categories is a folder named
according to the name of the module. This folder houses the previously
defined scripts. As an example, we will walk through the process of
creating a new module called \texttt{nonparametric}, as previously
mentioned in this section, which will perform a wilcoxon rank sum test.

Since the \texttt{nonparametric} module performs a statistical test, it
fits within the inference category, and hence should be placed in the
\texttt{intRo} repository at \textbf{modules/inference/nonparametric}.
Let's first create \emph{helper.R}:

\begin{Shaded}
\begin{Highlighting}[]
\NormalTok{nonparametrictest <-}\StringTok{ }\NormalTok{function(intro.data, x, y, }
                                \NormalTok{conflevel, althyp, hypval) \{}
    \KeywordTok{interpolate}\NormalTok{(~(}\KeywordTok{wilcox.test}\NormalTok{(}\DataTypeTok{x =} \NormalTok{df$x, }\DataTypeTok{y =} \NormalTok{df$y, }
                              \DataTypeTok{conf.level =} \NormalTok{conf, }
                              \DataTypeTok{alternative =} \NormalTok{althyp, }
                              \DataTypeTok{mu =} \NormalTok{hypval)),}
                  \DataTypeTok{df =} \KeywordTok{quote}\NormalTok{(intro.data),}
                  \DataTypeTok{x =} \NormalTok{x,}
                  \DataTypeTok{y =} \NormalTok{y,}
                  \DataTypeTok{conf =} \NormalTok{conflevel,}
                  \DataTypeTok{althyp =} \NormalTok{althyp,}
                  \DataTypeTok{hypval =} \NormalTok{hypval,}
                  \DataTypeTok{mydir =} \NormalTok{userdir, }
                  \StringTok{`}\DataTypeTok{_env}\StringTok{`} \NormalTok{=}\StringTok{ }\KeywordTok{environment}\NormalTok{(), }
                  \DataTypeTok{file =} \StringTok{"code_nonparametric.R"}\NormalTok{)}
\NormalTok{\}}
\end{Highlighting}
\end{Shaded}

This script is most immediately similar to standard R code. In this
case, a function \texttt{nonparametrictest} is created which, depending
on the values of the parameters, ultimately returns the result of a
wilcoxon rank sum test. One important difference from a typical R script
is that each call in the script is wrapped in a function called
\texttt{interpolate} (Wickham 2015). \texttt{interpolate} both executes
the given R code on the server, and also writes the code executed to the
script window at the bottom of \texttt{intRo}.

Because all the code needed to implement a wilcoxon rank sum test is
found in the \texttt{base} and \texttt{stats} package, the
\texttt{libraries.R} file will be empty for the \texttt{nonparametric}
module. Additionally, no reactive objects need be defined, so
\texttt{reactive.R} will also be empty. \texttt{observe.R}, which
defines the Shiny observers needed, can be written as follows:

\begin{Shaded}
\begin{Highlighting}[]
\KeywordTok{observe}\NormalTok{(\{}
    \KeywordTok{updateSelectizeInput}\NormalTok{(session, }\StringTok{"group1_non"}\NormalTok{, }
                         \DataTypeTok{choices =} \KeywordTok{intro.numericnames}\NormalTok{(), }
                         \DataTypeTok{selected =} \KeywordTok{ifelse}\NormalTok{(}\KeywordTok{checkVariable}\NormalTok{(}
                             \KeywordTok{intro.data}\NormalTok{(), input$group1_non), }
                             \NormalTok{input$group1_non, }
                             \KeywordTok{intro.numericnames}\NormalTok{()[}\DecValTok{1}\NormalTok{]))}
    \KeywordTok{updateSelectizeInput}\NormalTok{(session, }\StringTok{"group2_non"}\NormalTok{, }
                         \DataTypeTok{choices =} \KeywordTok{intro.numericnames}\NormalTok{(), }
                         \DataTypeTok{selected =} \KeywordTok{ifelse}\NormalTok{(}\KeywordTok{checkVariable}\NormalTok{(}
                             \KeywordTok{intro.data}\NormalTok{(), input$group2_non), }
                             \NormalTok{input$group2_non, }
                             \KeywordTok{intro.numericnames}\NormalTok{()[}\DecValTok{2}\NormalTok{]))}
\NormalTok{\})}

\KeywordTok{observeEvent}\NormalTok{(input$store_nonparametric, \{}
    \KeywordTok{cat}\NormalTok{(}\KeywordTok{paste0}\NormalTok{(}\StringTok{"}\CharTok{\textbackslash{}n\textbackslash{}n}\StringTok{"}\NormalTok{, }\KeywordTok{paste}\NormalTok{(}\KeywordTok{readLines}\NormalTok{(}
        \KeywordTok{file.path}\NormalTok{(userdir, }\StringTok{"code_nonparametric.R"}\NormalTok{)), }
        \DataTypeTok{collapse =} \StringTok{"}\CharTok{\textbackslash{}n}\StringTok{"}\NormalTok{)), }
        \DataTypeTok{file =} \KeywordTok{file.path}\NormalTok{(userdir, }\StringTok{"code_All.R"}\NormalTok{), }
        \DataTypeTok{append =} \OtherTok{TRUE}\NormalTok{)}
\NormalTok{\})}
\end{Highlighting}
\end{Shaded}

Shiny observers are a class of reactive objects within the Shiny
paradigm which do not return a value (RStudio and Inc. 2014). For
further discussion of reactivity, see Section
\ref{designing-for-reactivity}. In this example, observers are created
to ensure that the choices of variable for the \texttt{nonparametric}
module are only numeric variables. This is accomplished by utilizing the
global reactive \texttt{intro.numericnames()}, which returns a character
vector containing the variables in the current dataset that are numeric.
Finally, there is an event observer to store code generated from the
module into the overall code script upon clicking the store button. The
presence of this observer code and the definition of the button in the
user interface are enforced, and must be present in any \texttt{intRo}
module.

The \texttt{output.R} code can be very simple:

\begin{Shaded}
\begin{Highlighting}[]
\NormalTok{output$nonparametrictest <-}\StringTok{ }\KeywordTok{renderPrint}\NormalTok{(\{}
    \KeywordTok{return}\NormalTok{(}\KeywordTok{nonparametrictable}\NormalTok{(}\KeywordTok{intro.data}\NormalTok{(), input$group1_non, }
                              \NormalTok{input$group2_non, input$conflevel_non, }
                              \NormalTok{input$althyp_non, input$hypval_non))}
\NormalTok{\})}
\end{Highlighting}
\end{Shaded}

The \texttt{output.R} script then simply uses Shiny's
\texttt{renderPrint} function to display the resulting table.

Finally, a possible \texttt{ui.R} file is shown below:

\begin{Shaded}
\begin{Highlighting}[]
\NormalTok{nonparametric_ui <-}\StringTok{ }\KeywordTok{tabPanel}\NormalTok{(}\StringTok{"Nonparametric"}\NormalTok{,}
    \KeywordTok{column}\NormalTok{(}\DecValTok{4}\NormalTok{,}
         \KeywordTok{wellPanel}\NormalTok{(}
             \KeywordTok{selectizeInput}\NormalTok{(}\StringTok{"group1_non"}\NormalTok{, }\DataTypeTok{label =} \StringTok{"Group 1 (x)"}\NormalTok{, }
                            \DataTypeTok{choices =} \KeywordTok{numericNames}\NormalTok{(mpg), }
                            \DataTypeTok{selected =} \KeywordTok{numericNames}\NormalTok{(mpg)[}\DecValTok{1}\NormalTok{]),}
             \KeywordTok{selectizeInput}\NormalTok{(}\StringTok{"group2_non"}\NormalTok{, }\StringTok{"Group 2 (y)"}\NormalTok{, }
                            \DataTypeTok{choices =} \KeywordTok{numericNames}\NormalTok{(mpg), }
                            \DataTypeTok{selected =} \KeywordTok{numericNames}\NormalTok{(mpg)[}\DecValTok{2}\NormalTok{]),}
             
             \KeywordTok{hr}\NormalTok{(),}
             
             \KeywordTok{selectizeInput}\NormalTok{(}\StringTok{"althyp_non"}\NormalTok{, }\StringTok{"Alternative Hypothesis"}\NormalTok{, }
                            \KeywordTok{c}\NormalTok{(}\StringTok{"Two-Sided"} \NormalTok{=}\StringTok{ "two.sided"}\NormalTok{, }
                              \StringTok{"Greater"} \NormalTok{=}\StringTok{ "greater"}\NormalTok{, }\StringTok{"Less"} \NormalTok{=}\StringTok{ "less"}\NormalTok{)),}
             \KeywordTok{numericInput}\NormalTok{(}\StringTok{"hypval_non"}\NormalTok{, }\StringTok{"Hypothesized Value"}\NormalTok{,}
                          \DataTypeTok{value =} \DecValTok{0}\NormalTok{),}
             \KeywordTok{sliderInput}\NormalTok{(}\StringTok{"conflevel_non"}\NormalTok{, }\StringTok{"Confidence Level"}\NormalTok{,}
                         \DataTypeTok{min=}\FloatTok{0.01}\NormalTok{, }\DataTypeTok{max=}\FloatTok{0.99}\NormalTok{, }\DataTypeTok{step=}\FloatTok{0.01}\NormalTok{, }\DataTypeTok{value=}\FloatTok{0.95}\NormalTok{),}
             
             \KeywordTok{hr}\NormalTok{(),}
             
             \NormalTok{tags$}\KeywordTok{button}\NormalTok{(}\StringTok{""}\NormalTok{, }\DataTypeTok{id =} \StringTok{"store_nonparametric"}\NormalTok{, }\DataTypeTok{type =} \StringTok{"button"}\NormalTok{, }
                         \DataTypeTok{class =} \StringTok{"btn action-button"}\NormalTok{, }\KeywordTok{list}\NormalTok{(}\KeywordTok{icon}\NormalTok{(}\StringTok{"save"}\NormalTok{), }
                         \StringTok{"Store Nonparametric Result"}\NormalTok{), }
                         \DataTypeTok{onclick =} \StringTok{"$('#top-nav a:has(> .fa-print, }
\StringTok{                         .fa-code, .fa-download)').highlight();"}\NormalTok{)}
         \NormalTok{)}
    \NormalTok{),}
    
    \KeywordTok{column}\NormalTok{(}\DecValTok{8}\NormalTok{,}
         \NormalTok{tags$}\KeywordTok{b}\NormalTok{(}\StringTok{"Nonparametric Results"}\NormalTok{),}
         \KeywordTok{verbatimTextOutput}\NormalTok{(}\StringTok{"nonparametrictest"}\NormalTok{)}
    \NormalTok{)}
\NormalTok{)}
\end{Highlighting}
\end{Shaded}

This script defines all the inputs and outputs that the student will
see. The only requirements from \texttt{intRo}'s perspective are (1)
that there exist a store button at the bottom of the middle panel for
storing the results of the analysis in the code script, and (2) that
configuration options appear in the width 4 column in the middle, and
output appears in the width 8 column on the right. The remaining input
and output definitions depend on the statistical analysis or
transformation being performed.

Although the structure of an \texttt{intRo} module is relatively
straightforward, producing the code needed in a more seamless fashion
would certainly help open up the creation of such modules to a wider
audience. As we discuss in the conclusions and future work section,
providing an \texttt{intRo} module creation tool to abstract away some
of the less common coding paradigms, like the use of
\texttt{interpolate}, is an important effort that will continue to be
pursued.

\subsection{Designing for
Reproducibility}\label{designing-for-reproducibility}

While web-based tools written using Shiny (including \texttt{intRo})
have appealing characteristics such as being multi-platform, requiring
no installation, and requiring no software licenses, one limitation
immediately presents itself. The actions taken in the application are
typically not reproducible as in a standard R script. We have designed
\texttt{intRo} to overcome this limitation by capturing the unevaluated
expression of all actions taken by the user in the interface. This
expression is then parsed and printed in a code window at the bottom,
while simultaneously being executed by the R process running on the
server.

In essence, this procedure transcribes user actions in a GUI to R code.
When run in a standard R session, the results produced will be identical
to the results shown in \texttt{intRo}. The full series of actions taken
by the user are transcribed and can then be exchanged by researchers,
students, and developers in a manner similar to normal scripting. Even
``printing'' the results of an \texttt{intRo} session amounts to nothing
more than executing the given code on the server, and then storing the
results in an rmarkdown document, weaving the code with the results to
produce a full compendium. While not strictly necessary, this lends
credibility to the results produced by \texttt{intRo} in the sense that
they are directly reproduced by the server every time the user clicks
the print button.

Reproducibility in \texttt{intRo} is accomplished with the previously
mentioned \texttt{interpolate} function. \texttt{interpolate} accepts an
expression and an arbitrary number of arguments as an argument,
substitutes the arguments into the expression, prints the results to the
console, and evaluates the parsed expression. This allows for all
modules to be shoe-horned into the framework by wrapping the resulting R
code in calls to \texttt{interpolate}. A possible drawback of this
solution is that it requires module developers to manually wrap their
functions in this call, but this could be mitigated by a package that
creates modules automatically (See Section
\ref{conclusions-and-future-work}).

One potential enhancement to this framework would be the inclusion of
state-saving and state-resuming. Because an \texttt{intRo} session is
uniquely represented by the series of commands stored as code, the code
itself could represent a checkpoint for resuming a new \texttt{intRo}
session. Currently, beginning a new session will start the application
with no memory of previous sessions. In real-world usage, state saving
could allow a user to continue work later. At this time, this can only
be done by taking the code and pasting it into a standard R session,
although such an enhancement would likely involve minimal changes to
\texttt{intRo}'s underlying structure.s

\subsection{Designing for Reactivity}\label{designing-for-reactivity}

Reactive programming is a programming paradigm that ``tackles issues
posed by event-driven applications by providing abstractions to express
programs as reactions to external events and having the language
automatically manage the flow of time (by conceptually supporting
simultaneity), and data and computation dependencies'' (Bainomugisha et
al. 2012). As implemented by Shiny, results automatically update when
users interact with the interface.

intRo leverages the reactive programming nature of Shiny, and as such is
designed around the idea of user input cascading through the entire
application. In a typical Shiny application, users interact with inputs
that act as parameters to function, which in turn yield different
results. Within \texttt{intRo}, the students are able to interact with
and manipulate the data underlying the entire application. This posed
many challenges in the creation of \texttt{intRo} and drove design
decisions, namely timely save points according to the student's
workflow, and reactive updating of variable lists tied to inputs across
the entire application. Because the student may experiment with
different configurations or select different variables, we did not want
to store all actions taken in the intRo session. Rather, each module
includes a button allowing the student to explicitly store the output
visible in the results panel into the R script. This way, output is only
stored when the student is satisfied, and the resulting output is not
cluttered with unnecessary information.

In the creation of \texttt{intRo} we walked a fine line between giving
the student flexibility and having realistic usability. At the same
time, \texttt{intRo} was created as a consumer of another package,
Shiny, in which we as developers were the beneficiaries of another team
of developers' decision to balance flexibility and usability. For a
tangible example, consider the graphical summaries module. We only allow
variables of a type consistent with the selected plot to be displayed.
This is a conscious decision that limits an \texttt{intRo} user's
flexibility, while maximizing the usability (by minimizing crashes) of
the application. On the flip side of this, Shiny allows much higher
flexibility. For instance, the entire application (including user
interface) is created dynamically upon load, based on the modules
currently housed within \texttt{intRo}. However, Shiny does have limits
on its flexibility based on the designers decisions for usability. One
current example is the slider element. This element allows for fixed
width steps from its minimum to its maximum. The JavaScript library
being utilized in Shiny allows for arbitrary function calls to to
generate these steps, however they must be written in plain JavaScript.
This is an example of a decision made by the developers of Shiny to
limit functionality in favor of usability of their package.

\section{Conclusions and Future Work}\label{conclusions-and-future-work}

In this paper, we have outlined a framework for designing a web-based,
modular, extensible system which reproduces user actions into R code. We
believe that the development strategies we've outlined can and should be
applied to other software systems, as each of these characteristics aids
in the ease-of-use and functionality of the overall product. Although we
present them in the context of an introductory statistics application,
these ideas are generalizable and we hope that they will gain traction
in many other modern software systems.

With regards to \texttt{intRo} itself, we believe it can be a powerful
and effective tool for introductory statistics education. Its modular
structure allows it to be flexible enough for many different
applications and curriculums. Its ease-of-use allows the student to
focus her attention on the statistics task at hand, rather than
struggling with software licenses and confusing interface navigation.
Reproducible code generated from each analysis can be used to spark an
interest in R programming in those who might otherwise not be exposed to
it.

In addition to the current functionality, there are some practical
improvements in the works that will make \texttt{intRo} more useful to
both students and instructors. In particular, we have begun development
on an R package which will allow \texttt{intRo} modules to be created
automatically from user written R code. This package will generate the
necessary file structure to allow the module's incorporation into
\texttt{intRo} as well as translate user code to \texttt{intRo}
compatible code and populate the necessary files. This will vastly
improve \texttt{intRo}'s flexibility and allow it to be used in a wider
range of curricula, including more advanced statistics courses.
Additionally, we would like to expand the interactive capabilities of
our graphics in order to make \texttt{intRo}'s plots more engaging to
students. One way to do this would be implementing linked plots, in
which interactions with one plot are reflected in other plots that
illustrate the same data. This would be particularly useful in the
regression module so that students could explore observations with high
influence and leverage.

We hope to use \texttt{intRo} in courses to collect feedback regarding
the ease of use and functionality. This will allow us to assess its
usefulness relative to software used in the past, as well as gauge areas
for improvement. Furthermore, we can determine the effectiveness of code
printing on generating excitement from the students about programming in
R.

Challenges do exist with regards to the wider adoption of
\texttt{intRo}. For instance, we will need to monitor how well the
server hosting \texttt{intRo} handles the load of dozens of students
performing data analyses at once. If performance issues are encountered,
the infrastructure used may need to be expanded to handle current and
future load. An unknown quantity will be how feasible it is to increase
adoption of \texttt{intRo} across Iowa State, as well as to other
universities. One limitation of \texttt{intRo} is that uploading a
dataset beyond about 30,000 rows tends to be slow. Even once the data is
successfully uploaded, the default modules produce results more slowly
than with smaller datasets. This is a limitation that should be further
investigated if and when \texttt{intRo} sees wider adoption.

Regardless, tools that focus on usability and extensibility, such as
\texttt{intRo}, are sure to encourage the next round of innovators to be
interested and excited about statistical computing.

\section{Supplementary Material}\label{supplementary-material}

All code and documents related to this manuscript are available at
\url{https://github.com/gammarama/intRo}.

\section{Appendix}\label{appendix}

\subsection{Dynamic UI Generation}\label{dynamic-ui-generation}

\texttt{intRo}'s user interface and functionality is dynamically
generated depending on the set of modules enabled. The key driver to
populating \texttt{server.R} and \texttt{ui.R} is the modules folder,
the directory structure of which defines the placement of each module.
The interface is then created with the following statement.

\footnotesize

\begin{Shaded}
\begin{Highlighting}[]
\NormalTok{## Source ui}
\NormalTok{mylist <-}\StringTok{ }\KeywordTok{list}\NormalTok{()}
\NormalTok{old_heading <-}\StringTok{ ""}
\NormalTok{for (i in }\KeywordTok{seq_along}\NormalTok{(modules)) \{}
    \NormalTok{my.module <-}\StringTok{ }\KeywordTok{strsplit}\NormalTok{(modules[i], }\StringTok{"/"}\NormalTok{)[[}\DecValTok{1}\NormalTok{]]}
    \NormalTok{if (my.module[}\DecValTok{1}\NormalTok{] !=}\StringTok{ }\NormalTok{old_heading) \{}
        \NormalTok{mylist[[}\KeywordTok{length}\NormalTok{(mylist) +}\StringTok{ }\DecValTok{1}\NormalTok{]] <-}\StringTok{ }\NormalTok{Hmisc::}\KeywordTok{capitalize}\NormalTok{(my.module[}\DecValTok{1}\NormalTok{])}
        \NormalTok{old_heading <-}\StringTok{ }\NormalTok{my.module[}\DecValTok{1}\NormalTok{]}
    \NormalTok{\}}
    \NormalTok{mylist[[}\KeywordTok{length}\NormalTok{(mylist) +}\StringTok{ }\DecValTok{1}\NormalTok{]] <-}\StringTok{ }\KeywordTok{get}\NormalTok{(}\KeywordTok{paste}\NormalTok{(my.module[}\DecValTok{2}\NormalTok{], }
        \StringTok{"ui"}\NormalTok{, }\DataTypeTok{sep =} \StringTok{"_"}\NormalTok{))}
\NormalTok{\}}

\NormalTok{## mylist is a list containing the different ui}
\NormalTok{## module code Create the UI}
\KeywordTok{shinyUI}\NormalTok{(}\KeywordTok{navbarPage}\NormalTok{(}\StringTok{"intRo"}\NormalTok{, }\DataTypeTok{id =} \StringTok{"top-nav"}\NormalTok{, }\DataTypeTok{theme =} \StringTok{"bootstrap.min.css"}\NormalTok{, }
    \KeywordTok{tabPanel}\NormalTok{(}\DataTypeTok{title =} \StringTok{""}\NormalTok{, }\DataTypeTok{icon =} \KeywordTok{icon}\NormalTok{(}\StringTok{"home"}\NormalTok{), }\KeywordTok{fluidRow}\NormalTok{(}\KeywordTok{do.call}\NormalTok{(navlistPanel, }
        \KeywordTok{c}\NormalTok{(}\KeywordTok{list}\NormalTok{(}\DataTypeTok{id =} \StringTok{"side-nav"}\NormalTok{, }\DataTypeTok{widths =} \KeywordTok{c}\NormalTok{(}\DecValTok{2}\NormalTok{, }\DecValTok{10}\NormalTok{)), }
            \NormalTok{mylist)))), ...))}
\end{Highlighting}
\end{Shaded}

\normalsize

The key piece of code being the \texttt{do.call} statement loading the
list of ui elements from the module's \texttt{ui.R} file. The server
functions are then dynamically generated using a similar method.

\footnotesize

\begin{Shaded}
\begin{Highlighting}[]
\KeywordTok{shinyServer}\NormalTok{(function(input, output, session) \{}
    \NormalTok{types <-}\StringTok{ }\KeywordTok{c}\NormalTok{(}\StringTok{"helper.R"}\NormalTok{, }\StringTok{"observe.R"}\NormalTok{, }\StringTok{"reactive.R"}\NormalTok{, }
        \StringTok{"output.R"}\NormalTok{)}
    
    \NormalTok{modules_tosource <-}\StringTok{ }\KeywordTok{file.path}\NormalTok{(}\StringTok{"modules"}\NormalTok{, }\KeywordTok{apply}\NormalTok{(}\KeywordTok{expand.grid}\NormalTok{(modules, }
        \NormalTok{types), }\DecValTok{1}\NormalTok{, paste, }\DataTypeTok{collapse =} \StringTok{"/"}\NormalTok{))}
    
    \NormalTok{for (mod in modules_tosource) \{}
        \KeywordTok{source}\NormalTok{(mod, }\DataTypeTok{local =} \OtherTok{TRUE}\NormalTok{)}
    \NormalTok{\}}
\NormalTok{\})}
\end{Highlighting}
\end{Shaded}

\normalsize
In this way, we were able to have \texttt{intRo} be fully extensible,
its structure and functionality dependent entirely on the modules
present within the application.

\subsection{\texorpdfstring{Deploying \texttt{intRo}
Instances}{Deploying intRo Instances}}\label{deploying-intro-instances}

Although students can access intRo from
\url{http://www.intro-stats.com}, course instructors may wish to
download, customize, and deploy their own instance, perhaps with new
modules or modified theming or functionality. \texttt{intRo} can be
downloaded, ran, and deployed on ShinyApps.io through the use of the R
package \texttt{intRo}. Currently, the package is only available on
GitHub, and can be installed using the devtools package as follows:

\begin{Shaded}
\begin{Highlighting}[]
\NormalTok{devtools::}\KeywordTok{install_github}\NormalTok{(}\StringTok{"gammarama/intRo"}\NormalTok{)}
\end{Highlighting}
\end{Shaded}

After installing the \texttt{intRo} package, the first function one
should call is \texttt{download\_intRo}. \texttt{download\_intRo} takes
as an argument a directory in which to store the application. By
default, it selects the working directory of the R session. This
function clones the application branch of the \texttt{intRo} repository
on GitHub, and hence will pull the latest version of the code whenever
it is ran.

Running \texttt{download\_intRo} will produce an \texttt{intRo} folder
in the specified folder. It can then be ran as any Shiny application,
using Shiny's \texttt{runApp} command. However, we have provided a
wrapper function \texttt{run\_intRo} which adds some additional
customization options to the execution process. \texttt{run\_intRo}
takes as argument the path to the folder containing the \texttt{intRo}
application. It also takes several more optional arguments:

\begin{itemize}
\tightlist
\item
  \texttt{enabled\_modules}: A character vector containing the modules
  to enable
\item
  \texttt{theme}: A string representing a shinythemes theme to use
\item
  \texttt{...}: Additional arguments passed to Shiny's \texttt{runApp}
  function
\end{itemize}

The package provides help documentation which explains in further detail
the format that these arguments would take, but as an example, suppose I
wanted to download \texttt{intRo} to my working directory, execute an
\texttt{intRo} session with only the data sources, data transform, and
numerical summaries modules enabled, and apply the cerulean theme. The
series of calls to do so would be as follows:

\begin{Shaded}
\begin{Highlighting}[]
\KeywordTok{download_intRo}\NormalTok{()}
\KeywordTok{run_intRo}\NormalTok{(}\DataTypeTok{enabled_modules =} \KeywordTok{c}\NormalTok{(}\StringTok{"data/transform"}\NormalTok{, }\StringTok{"summaries/numerical"}\NormalTok{), }
          \DataTypeTok{theme =} \StringTok{"cerulean"}\NormalTok{)}
\end{Highlighting}
\end{Shaded}

Note that the data sources module is required, and hence must be
included in all intRo sessions and need not be specified in the
enabled\_modules argument.

If the intent is to use a specific instance of \texttt{intRo} where many
students will access it at the same time, such as in an introductory
statistics class, it may be preferable to deploy a custom instance of
\texttt{intRo} to a publicly accessible URL. The package provides a
function \texttt{deploy\_intRo} which is a wrapper for the
\texttt{deployApp} function contained in the shinyapps package. Once the
shinyapps package is installed and configured, \texttt{deploy\_intRo}
will upload \texttt{intRo} as an application on the instructor's
ShinyApps.io account. The function takes the same arguments as
\texttt{run\_intRo}, so it can be deployed with a custom selection of
modules, and a customized theme. It also takes an additional argument
\texttt{google\_analytics}, which allows the specification of a Google
Analytics tracking ID. It also takes \texttt{...} as additional
arguments to be passed into the \texttt{deployApp} routine. For example,
if we wished to deploy the instance of \texttt{intRo} we ran previously,
we would call it like so:

\begin{Shaded}
\begin{Highlighting}[]
\KeywordTok{deploy_intRo}\NormalTok{(}\DataTypeTok{enabled_modules =} \KeywordTok{c}\NormalTok{(}\StringTok{"data/transform"}\NormalTok{, }\StringTok{"summaries/numerical"}\NormalTok{), }
             \DataTypeTok{theme =} \StringTok{"cerulean"}\NormalTok{)}
\end{Highlighting}
\end{Shaded}

Once the process finished, the app will become available at
\url{http://<user>.shinyapps.io/intRo}, where
\texttt{\textless{}user\textgreater{}} is the username of the
ShinyApps.io account configured.

\section*{References}\label{references}
\addcontentsline{toc}{section}{References}

\hypertarget{refs}{}
\hypertarget{ref-rmarkdown}{}
Allaire, JJ, Jonathan McPherson, Yihui Xie, Hadley Wickham, Joe Cheng,
and Jeff Allen. 2014. \emph{Rmarkdown: Dynamic Documents for R}.
\url{http://CRAN.R-project.org/package=rmarkdown}.

\hypertarget{ref-baggerly2011reproducible}{}
Baggerly, Keith A, and Donald A Berry. 2011. ``Reproducible Research.''
\emph{AMSTAT News: The Membership Magazine of the American Statistical
Association}, no. 403. American Statistical Association: 16--17.

\hypertarget{ref-bainomugisha2012survey}{}
Bainomugisha, Engineer, Andoni Lombide Carreton, Tom Van Cutsem, Stijn
Mostinckx, and Wolfgang De Meuter. 2012. ``A Survey on Reactive
Programming.'' In \emph{ACM Computing Surveys}. Citeseer.

\hypertarget{ref-swirl}{}
Carchedi, Nick, Bill Bauer, Gina Grdina, and Sean Kross. 2014.
\emph{Swirl: Learn R, in R.}
\url{http://CRAN.R-project.org/package=swirl}.

\hypertarget{ref-shinymodules}{}
Cheng, Joe. 2015. ``Shiny - Modularizing Shiny App Code.''
\url{http://shiny.rstudio.com/articles/modules.html}.

\hypertarget{ref-datacamp}{}
DataCamp. 2014. ``Online R Tutorials and Data Science Courses -
Datacamp.'' \url{https://www.datacamp.com/}.

\hypertarget{ref-fellows2012}{}
Fellows, Ian. 2012. ``Deducer: A Data Analysis Gui for R.''
\emph{Journal of Statistical Software} 49 (8).

\hypertarget{ref-fox2005}{}
Fox, John. 2005. ``The R Commander: A Basic-Statistics Graphical User
Interface to R.'' \emph{Journal of Statistical Software} 14 (9).

\hypertarget{ref-radiant}{}
Nijs, Vincent. 2016. ``Radiant - Business Analytics Using R and Shiny.''
\url{https://radiant-rstats.github.io/docs}.

\hypertarget{ref-parnas1972criteria}{}
Parnas, David Lorge. 1972. ``On the Criteria to Be Used in Decomposing
Systems into Modules.'' \emph{Communications of the ACM} 15 (12). ACm:
1053--8.

\hypertarget{ref-mosaic}{}
Pruim, Randall, Daniel Kaplan, and Nicholas Horton. 2014. \emph{Mosaic:
Project Mosaic (Mosaic-Web.org) Statistics and Mathematics Teaching
Utilities}. \url{http://CRAN.R-project.org/package=mosaic}.

\hypertarget{ref-r-stat}{}
R Core Team. 2014. \emph{R: A Language and Environment for Statistical
Computing}. Vienna, Austria: R Foundation for Statistical Computing.
\url{http://www.R-project.org/}.

\hypertarget{ref-shiny}{}
RStudio, and Inc. 2014. \emph{Shiny: Web Application Framework for R}.
\url{http://CRAN.R-project.org/package=shiny}.

\hypertarget{ref-scrimshaw2001computers}{}
Scrimshaw, Peter. 2001. ``Computers and the Teacher's Role.''
\emph{Knowledge, Power and Learning}. London, Paul Chapman Publishing
Ltd.

\hypertarget{ref-szyperski1996independently}{}
Szyperski, Clemens. 1996. ``Independently Extensible Systems-Software
Engineering Potential and Challenges.'' \emph{Australian Computer
Science Communications} 18. UNIVERSITY OF CANTERBURY: 203--12.

\hypertarget{ref-5359977}{}
Tan, P. H., C. Y. Ting, and S. W. Ling. 2009. ``Learning Difficulties in
Programming Courses: Undergraduates' Perspective and Perception.'' In
\emph{Computer Technology and Development, 2009. Icctd '09.
International Conference on}, 1:42--46.
doi:\href{https://doi.org/10.1109/ICCTD.2009.188}{10.1109/ICCTD.2009.188}.

\hypertarget{ref-interpolate}{}
Wickham, Hadley. 2015. ``Graphics \& Computing Student Paper Winners @
Jsm 2015.'' \url{https://github.com/hadley/15-student-papers}.

\hypertarget{ref-inzight}{}
Wild, Chris. 2015. ``INZight Lite.''
\url{http://lite.docker.stat.auckland.ac.nz}.

\hypertarget{ref-xie2015}{}
Xie, Yihui. 2015. \emph{Dynamic Documents with R and Knitr}. Vol. 29.
CRC Press.

\end{document}